\documentstyle[12pt]{article}                    %%%
\newfont{\ffont}{msym10}                        %%
\newcommand{\beq}{\begin{equation}}             %%
\newcommand{\eeq}{\end{equation}}               %%
\newcommand{\bqry}{\begin{eqnarray}}            %%
\newcommand{\eqry}{\end{eqnarray}}              %%
\newcommand{\bqryn}{\begin{eqnarray*}}          %%
\newcommand{\eqryn}{\end{eqnarray*}}            %%
\newcommand{\preprint}[1]{\begin{table}[t]      %%
            \begin{flushright}                  %%
            \begin{large}{#1}\end{large}        %%
            \end{flushright}                    %%
            \end{table}}                        %%
\newcommand{\PD}[2]                             %%
    {\frac{\partial^{#2}}{\partial #1^{#2}}}    %%
\begin{document}
\preprint{LA-UR-97-XXXX \\ TAUP-XXXX-97}
\title{New Mass Relation \\ for Meson 25-plet}
\author{\\ L. Burakovsky\thanks{E-mail: BURAKOV@PION.LANL.GOV}, \
T. Goldman\thanks{E-mail: GOLDMAN@T5.LANL.GOV} \
\\  \\  Theoretical Division, MS B285 \\  Los Alamos National  
Laboratory \\
Los Alamos, NM 87545, USA \\  \\  and  \\  \\
L.P. Horwitz\thanks{E-mail: HORWITZ@TAUNIVM.TAU.AC.IL. Also at  
Department of
Physics, Bar-Ilan University, Ramat-Gan, Israel  } \
\\  \\ School of Physics and Astronomy \\ Tel-Aviv University \\  
Ramat-Aviv,
69978 Israel \\}
\date{ }
\maketitle
\begin{abstract}
By assuming the existence of (quasi)-linear Regge trajectories for  
25-plet
mesons in the low energy region, we derive a new, 14th power, meson 
mass relation. This relation may be reduced to a quadratic  
Gell-Mann--Okubo
type formula by fitting the values of the Regge slopes of these  
(quasi)-linear
trajectories. Such a formula holds with an accuracy of $\sim 2$\%  
for vector
mesons, and also suggests that the quantum number  
$I(J^P)=\frac{1}{2}(2^{+})$
should be assigned to the both $B_J(5732)$ and $B_{sJ}(5850)$ mesons 
discovered recently. We also discuss reasons for the failure of  
group theory
to produce correct mass relations for heavy quarkonia.
\end{abstract}

\bigskip

\section*{  }
The generalization of the standard $SU(3)$ Gell-Mann--Okubo mass formula 
\cite{GMO} to higher symmetry groups, e.g., $SU(4)$ and $SU(5),$  
became a
natural subject of investigation after the discovery of the fourth  
and fifth
quark flavors in the mid-70's \cite{disc}. Attempts have been made  
in the
literature to derive such a formula, either quadratic or linear in  
mass, by a)
using group theoretical methods \cite{Mac,BMO,Boal,Iwao,DMP}, b)  
generalizing
the perturbative treatment of $U(3)\times U(3)$ chiral symmetry  
breaking and
the corresponding Gell-Mann-Oakes-Renner relation \cite{GOR} to  
$U(4)\times
U(4)$ \cite{GLR,MRT}, c) assuming the asymptotic realization of $SU(4)$ 
symmetry in the algebra $[A_\alpha ,A_\beta ]=if_{\alpha \beta  
\gamma }V_{
\gamma }$ (where $V_{\alpha },A_{\beta }$ are vector and  
axial-vector charges,
respectively) \cite{HO}, d) extending the Weinberg spectral  
function sum rules
\cite{Wei} to accommodate the higher symmetry breaking effects  
\cite{BGN}, and
e) applying alternative methods, such as the linear mass spectrum  
for meson
multiplets\footnote{Here, we speak of linear spectrum over the multiplet 
quantum numbers, taking proper account of degeneracy, not  
(directly) make use
of linear Regge trajectories.} \cite{lin,su4}. In the  
following\footnote{Here
$\eta $ stands for the masses of both isovector and isoscalar $n\bar{n}$ 
states which coincide on a naive quark model level.}, $\eta ,\eta  
_s,\eta _c,
\eta _b,K,D,D_s,B,B_s,B_c$ stand for the masses of the $n\bar{n}$  
$(n\equiv u$
or  
$d),s\bar{s},c\bar{c},b\bar{b},s\bar{n},c\bar{n},c\bar{s},b\bar{n},b\bar{s},
b\bar{c}$ mesons, respectively\footnote{Since these designations  
apply to all
spin states, vector mesons will be confusingly labelled as $\eta $'s.
We ask the reader to bear with us in this in the interest of minimizing 
notation.} The linear mass relations
\bqry
D & = & \frac{\eta +\eta _c}{2},\;\;\;D_s\;=\;\frac{\eta _s+\eta  
_c}{2}, \\
B & = & \frac{\eta +\eta _b}{2},\;\;\;B_s\;=\;\frac{\eta _s+\eta  
_b}{2},\;\;\;
B_c\;=\;\frac{\eta _c+\eta _b}{2}
\eqry
found in \cite{Boal,Iwao,MRT}, although perhaps justified for  
vector mesons,
since a vector meson mass is given approximately by a sum of the  
corresponding
constituent quark masses, $$m(i\bar{j})\simeq m(i)+m(j)$$ (in fact,  
for vector
mesons, the relations (1),(2) hold with an accuracy of up to $\sim 4$\%),
are expected to fail for other meson multiplets, as confirmed by direct
comparison with experiment. Similarly, the quadratic mass relations
\beq
B_s^2-B^2=D_s^2-D^2=K^2-\eta ^2
\eeq
obtained in ref. \cite{GLR} by generalizing the $SU(3)$  
Gell-Mann-Oakes-Renner
relation \cite{GOR} to include the $D$ and $D_s$ mesons (here $\pi  
$ stands
for the mass of the $\pi ,$ etc.),
\beq
\frac{\pi ^2}{2n}=\frac{K^2}{n+s}=\frac{D^2}{n+c}=\frac{D_s^2}{s+c},
\eeq
(and therefore $D_s^2-D^2=K^2-\pi ^2\propto (s-n),$ also found in refs. 
\cite{BMO,HO}), does not agree with experiment. For pseudoscalar mesons,
for example, one has (in GeV$^2)$ 0.388 for $(D_s^2-D^2)$ vs. 0.226  
for $(K^2-
\pi ^2).$ For vector mesons, the corresponding quantities are  0.424 vs. 
0.199, with about 100\% discrepancy. The reason that the relations  
(3) do not
hold is apparently due to the impossibility of perturbative  
treatment of $U(4)
\times U(4)$ and $U(5)\times U(5)$ symmetry breaking, as a  
generalization of
that of $U(3)\times U(3),$ due to very large bare masses of the $c$- and 
$b$-quarks as compared to those of the $u$-, $d$- and $s$-quarks.  
In ref.
\cite{su4} by the application of the linear spectrum to $SU(4)$  
meson 16-plet,
the following relation was obtained,
\beq
12\bar{D}^2=7\eta _0^2+5\eta_c^2,
\eeq
where $\bar{D}$ is the average mass of the $D$ and $D_s$ states  
(which are mass
degenerate when $SU(4)$ flavor symmetry is broken to $SU(3)),$ and  
$\eta _0$
is the mass average of the corresponding $SU(3)$ nonet. As shown in ref. 
\cite{su4}, this relation holds with an accuracy of up to $\sim $  
5\% for all
well established meson 16-plets. The accuracy of the generalization  
of Eq.
(5) to a meson 25-plet is, however, much worse than that of (5),  
perhaps due
to the contribution of the higher power corrections to the linear  
spectrum in
the mass range corresponding to the $b$-flavored mesons \cite{su4}.    

It is well known that the hadrons composed of light $(u,d,s)$  
quarks populate
linear Regge trajectories; i.e., the square of the mass of a state with 
orbital momentum $\ell $ is proportional to $\ell :$ $M^2(\ell  
)=\ell /\alpha
^{'}+\;{\rm const,}$ where the slope $\alpha ^{'}$ depends weakly on the 
flavor content of the states lying on the corresponding trajectory,
\beq
\alpha ^{'}_{n\bar{n}}\simeq 0.88\;{\rm GeV}^{-2},\;\;\;
\alpha ^{'}_{s\bar{n}}\simeq 0.84\;{\rm GeV}^{-2},\;\;\;
\alpha ^{'}_{s\bar{s}}\simeq 0.80\;{\rm GeV}^{-2}.
\eeq
In contrast, the data on the properties of Regge trajectories of hadrons 
containing heavy quarks are almost nonexistent at the present time,  
although
it is established \cite{BB} that the slope of the trajectories  
decreases with
increasing quark mass (as seen in Eq. (6)) in the mass region of  
the lowest
excitations. This is due to an increasing (with mass) contribution  
of the
color Coulomb interaction, leading to a curvature of the trajectory  
near the
ground state. However, as the analyses show \cite{BB,KS,QR}, in the  
asymptotic
regime of the highest excitations, the trajectories of both light  
and heavy
quarkonia are linear and have the same slope $\alpha ^{'}\simeq  
0.9$ GeV$^{-2
},$ in agreement with natural expectations from the string model.

Knowledge of Regge trajectories in the scattering region, i.e., at  
$t<0,$ and
of the intercepts $a(0)$ and slopes $\alpha ^{'}$ is also useful  
for many
non-spectral purposes, for example, in the recombination \cite{rec} and 
fragmentation \cite{fra} models. Therefore, as pointed out in ref.  
\cite{BB},
the slopes and intercepts of the Regge trajectories are the fundamental 
constants of hadron dynamics, perhaps generally more important than  
the mass
of any particular state. Thus, not only the derivation of a mass  
relation but
also the determination of the parameters $a(0)$ and $\alpha ^{'}$  
of heavy
quarkonia is of great importance, since they afford opportunities  
for better
understanding of the dynamics of the strong interactions in the  
processes of
production of charmed and beauty hadrons at high energies.

Here we apply Regge phenomenology for the derivation of a mass  
relation for the
$SU(5)$ meson 25-plet, by assuming the (quasi)-linear form of Regge 
trajectories for heavy quarkonia with slopes which are generally  
different
from (less than) the standard one, $\alpha ^{'}\simeq 0.9$  
GeV$^{-2}.$ We show
that for this relation to avoid depending on the values of the  
slopes, it
must be of 14th power in meson masses. The relation may be reduced to a 
quadratic Gell-Mann--Okubo type formula by fitting the values of  
the slopes.
We note that the corresponding formula for the $SU(4)$ meson  
16-plet, which is
of sixth power in meson masses, was derived in our previous paper  
\cite{su41}.

Let us assume the (quasi)-linear form of Regge trajectories for  
hadrons with
identical $J^{PC}$ quantum numbers (i.e., belonging to a common  
multiplet).
Then for the states with orbital momentum $\ell $ one has
\bqryn
\ell & = & \alpha ^{'}_{i\bar{i}}m^2_{i\bar{i}}\;+a_{i\bar{i}}(0), \\    
\ell & = & \alpha  
^{'}_{j\bar{i}}m^2_{j\bar{i}}\;\!+a_{j\bar{i}}(0), \\
\ell & = & \alpha ^{'}_{j\bar{j}}m^2_{j\bar{j}}+a_{j\bar{j}}(0).    
\eqryn
Using now the relation among the intercepts \cite{inter,Kai},
\beq
a_{i\bar{i}}(0)+a_{j\bar{j}}(0)=2a_{j\bar{i}}(0),
\eeq
one obtains from the above relations
\beq
\alpha ^{'}_{i\bar{i}}m^2_{i\bar{i}}+\alpha  
^{'}_{j\bar{j}}m^2_{j\bar{j}}=
2\alpha ^{'}_{j\bar{i}}m^2_{j\bar{i}}.
\eeq
In order to eliminate the Regge slopes from this formula, we need a  
relation
among the slopes. Two such relations have been proposed in the  
literature,
\beq
\alpha ^{'}_{i\bar{i}}\cdot \alpha ^{'}_{j\bar{j}}=\left( \alpha  
^{'}_{j\bar{i
}}\right) ^2,
\eeq
which follows from the factorization of residues of the $t$-channel poles
\cite{first,KY}, and
\beq
\frac{1}{\alpha ^{'}_{i\bar{i}}}+\frac{1}{\alpha  
^{'}_{j\bar{j}}}=\frac{2}{
\alpha ^{'}_{j\bar{i}}},
\eeq
based on topological expansion and the $q\bar{q}$-string picture of  
hadrons
\cite{Kai}.

For light quarkonia (and small differences in the $\alpha ^{'}$  
values), there
is no essential difference between these two relations; viz., for  
$\alpha ^{'
}_{j\bar{i}}=\alpha ^{'}_{i\bar{i}}/(1+x),$ $x\ll 1,$ Eq. (10)  
gives $\alpha
^{'}_{j\bar{j}}=\alpha ^{'}_{i\bar{i}}/(1+2x),$ whereas Eq. (9)  
gives $\alpha
^{'}_{j\bar{j}}=\alpha ^{'}_{i\bar{i}}/(1+x)^2\approx \alpha  
^{'}/(1+2x),$ i.e,
differing only in the interpretation of the value of $x,$ to order  
$x^2.$
However, for heavy quarkonia (and expected large differences from  
the $\alpha
^{'}$ values for the light quarkonia) these relations are  
incompatible; e.g.,
for $\alpha ^{'}_{j\bar{i}}=\alpha ^{'}_{i\bar{i}}/2,$ Eq. (9) will give 
$\alpha ^{'}_{j\bar{j}}=\alpha ^{'}_{i\bar{i}}/4,$ whereas Eq. (10)  
produces
$\alpha ^{'}_{j\bar{j}}=\alpha ^{'}_{i\bar{i}}/3.$ One has  
therefore to choose
one of these relations in order to proceed
further. Here we use Eq. (10), since it is much more consistent  
with (8) than
is Eq. (9), which we tested by using measured quarkonia masses in  
Eq. (8). We
shall justify this choice in more detail in a separate publication  
\cite{prep}.

Since we are interested in $SU(4)$ and $SU(5)$ breaking, and since it 
simplifies the discussion, we take average slope in the light quark  
sector:
\beq
\alpha ^{'}_{n\bar{n}}\cong \alpha ^{'}_{s\bar{n}}\cong \alpha  
^{'}_{s\bar{s}}
\cong \alpha ^{'}\simeq 0.85\;{\rm GeV}^{-2}.
\eeq
Note that this should lead us to expect accuracy to be limited to  
$\sim 5$\%
in what follows.

It then follows from the relations based on (8),
\bqry
\alpha ^{'}\eta ^2+\alpha ^{'}_{c\bar{c}}\eta _c^2 & = & 2\alpha  
^{'}_{c\bar{n
}}D^2, \\
\alpha ^{'}\eta _s^2+\alpha ^{'}_{c\bar{c}}\eta _c^2 & = & 2\alpha  
^{'}_{c
\bar{s}}D_s^2,
\eqry
that
\beq
\alpha ^{'}_{c\bar{n}}\cong \alpha ^{'}_{c\bar{s}}=\alpha  
^{'}\frac{(\eta _s^2-
\eta ^2)}{2(D_s^2-D^2)},
\eeq
\beq
\alpha ^{'}_{c\bar{c}}=\alpha ^{'}\left[ \frac{(\eta _s^2-\eta  
^2)}{(D_s^2-D^
2)}\frac{D^2}{\eta _c^2}-\frac{\eta ^2}{\eta _c^2}\right] .
\eeq
Using these values of the slopes in Eq. (10) with $i=n,\;j=c$, we obtain
\beq
\left( \eta _s^2D^2 - \eta ^2D_s^2\right) \left( \eta  
_s^2-\eta^2\right) +
\eta _c^2\left( D_s^2-D^2\right) \left( \eta _s^2-\eta ^2\right) =4\left(
\eta _s^2D^2-\eta ^2D_s^2\right) \left( D_s^2-D^2\right) ,
\eeq
which is the new mass relation for the $SU(4)$ meson 16-plet  
obtained in our
previous paper \cite{su41}. As shown in \cite{su41}, this relation  
holds with
an accuracy of up to $\sim 5$\% for the four well-established meson  
multiplets.
The use of $B,B_s,B_c$ in place of $D,D_s,D_c,$ respectively, in  
Eq. (12),(13)
leads to
\beq
\alpha ^{'}_{b\bar{n}}\cong \alpha ^{'}_{b\bar{s}}=\alpha  
^{'}\frac{(\eta _s^2-
\eta ^2)}{2(B_s^2-B^2)},
\eeq
\beq
\alpha ^{'}_{b\bar{b}}=\alpha ^{'}\left[ \frac{(\eta _s^2-\eta  
^2)}{(B_s^2-B^
2)}\frac{B^2}{\eta _b^2}-\frac{\eta ^2}{\eta _b^2}\right] .
\eeq
Using now these values of the slopes in Eq. (10) with $i=n,\;j=b$,  
we obtain
\beq
\left( \eta _s^2B^2 - \eta ^2B_s^2\right) \left( \eta  
_s^2-\eta^2\right) +
\eta _b^2\left( B_s^2-B^2\right) \left( \eta _s^2-\eta ^2\right) =4\left(
\eta _s^2B^2-\eta ^2B_s^2\right) \left( B_s^2-B^2\right) ,
\eeq
which is a new mass relation for the $SU(4)$ meson 16-plet built of  
the $u$-,
$d$-, $s$-, and $b$-quarks \cite{su41}. At present, this relation  
may only be
tested for vector mesons alone since the masses of all of the  
beauty states
involved are established experimentally only for vector mesons  
\cite{data}. As
shown in \cite{su41}, in this case the accuracy is $\sim 3.5$\%.

In order to extend the results to $SU(5)$ and to incorporate the  
remaining
$b\bar{c}$-state of the 25-plet, we use a relation based on (8),
\beq
\alpha ^{'}_{c\bar{c}}\eta _c^2+\alpha ^{'}_{b\bar{b}}\eta  
_b^2=2\alpha ^{'}_{
b\bar{c}}B_c^2,
\eeq
which leads, through (15),(18), to
\beq
\alpha ^{'}_{b\bar{c}}=\alpha ^{'}\left[ \frac{\eta _s^2-\eta  
^2}{B_s^2-B^2}
\frac{B^2}{2B_c^2}+\frac{\eta _s^2-\eta ^2}{D_s^2-D^2}\frac{D^2}{2B_c^2}-
\frac{\eta ^2}{B_c^2}\right] .
\eeq
Using now this expression for $\alpha ^{'}_{b\bar{c}}$ together  
with those
for $\alpha ^{'}_{c\bar{c}}$ and $\alpha ^{'}_{b\bar{b}},$ Eqs.  
(15),(18), in
relation, as follows from (10), $$\frac{1}{\alpha ^{'}_{b\bar{b}}}+\frac{
1}{\alpha ^{'}_{c\bar{c}}}=\frac{2}{\alpha ^{'}_{b\bar{c}}},$$ we finally
arrive at
$$\left[ (\eta _s^2-\eta ^2)\left( \eta _c^2B^2(D_s^2-D^2)+\eta  
_b^2D^2(B_s^2-
B^2)\right) -\eta ^2(\eta _c^2+\eta  
_b^2)(D_s^2-D^2)(B_s^2-B^2)\right] $$
$$\times \left[ (\eta _s^2-\eta ^2)\left(  
B^2(D_s^2-D^2)+D^2(B_s^2-B^2)\right)
-2\eta ^2(D_s^2-D^2)(B_s^2-B^2)\right] $$
\beq
=4B_c^2(D_s^2-D^2)(B_s^2-B^2)\left[ D^2(\eta _s^2-\eta ^2)-\eta  
^2(D_s^2-D^2)
\right] \left[ B^2(\eta _s^2-\eta ^2)-\eta ^2(B_s^2-B^2)\right] ,
\eeq
which is a new, 14th power in meson masses, relation for the  
$SU(5)$ 25-plet.
As applied to vector mesons, this relation yields
for the mass of the $B_c^\ast :$
\beq
m(B_c)=6.285\;{\rm GeV},
\eeq
in agreement with a rough estimate, as follows from (2),
\beq
m(B_c^\ast )\simeq \frac{m(J/\psi )+m(\Upsilon (1S))}{2}\cong  
6.280\;{\rm GeV.}
\eeq

We emphasize that the formulas (16),(19),(22) do not depend on the  
$values$ of
the Regge slopes, but only on a relation between them, Eq. (10), which 
justifies their use in both the low energy region where the slopes are 
different and the high energy region where all the slopes coincide.  
In the
latter case, for example, as follows from (12),(13), $\eta  
_s^2-\eta ^2=2(D_s^
2-D^2),$ and Eq. (16) reduces to
\beq
\eta ^2+\eta _c^2=2D^2,
\eeq
consistent with Eq. (12) in this limit. One may also find from Eqs. 
(12),(13) and similar relations for the beauty mesons with equal  
slopes, and
the standard $SU(3)$ Gell-Mann--Okubo relation,
\beq
\eta ^2+\eta _s^2=2K^2,
\eeq
that the relations (3) also hold in this limit.

Let us now discuss the question of the generalization of the standard 
$SU(3)$ Gell-Mann--Okubo mass formula which is quadratic in mass to  
the case
of heavier quarkonia. We shall continue to assume the validity of  
Eq. (11)
and introduce $x>0$ through the relation
\beq
\alpha ^{'}_{c\bar{n}}=\alpha ^{'}_{c\bar{s}}=\frac{\alpha ^{'}}{1+x}.
\eeq
It then follows from (10) that
\beq
\alpha ^{'}_{c\bar{c}}=\frac{\alpha ^{'}}{1+2x},
\eeq
and one obtains from (12),(13),
\beq
(1+x)\left( \eta ^2+\eta _s^2\right) +\frac{2(1+x)}{1+2x}\eta  
_c^2=2\left( D^2
+D_s^2\right) .
\eeq
Results of the calculations of the Regge slopes of heavy quarkonia  
in refs.
\cite{KY}: $\alpha ^{'}_{c\bar{n}}/\alpha ^{'}\simeq \alpha  
^{'}_{c\bar{s}}/
\alpha ^{'}\simeq 0.73,$ $\alpha ^{'}_{c\bar{c}}/\alpha ^{'}\simeq  
0.58,$ and
\cite{Kai,CN}: $\alpha ^{'}_{c\bar{c}}\simeq 0.5$ GeV$^{-2},$  
support the value
\beq
x\cong 0.355.
\eeq
With this $x,$ it follows from (29) and the standard $SU(3)$  
Gell-Mann--Okubo
formula (19) that
\beq
8.13\;K^2+4.75\;\eta _c^2=6\left( D^2+D_s^2\right) .
\eeq
As shown in ref. \cite{su41}, this formula holds at a 1\% level for  
the four
well-established meson multiplets. Also, the formula (31) is in  
qualitative
agreement with the relation (5) obtained by two of the present  
authors in ref.
\cite{su4} by the application of the linear spectrum to a meson 16-plet.

The entire analysis may, of course, be repeated for $B,B_s,\eta _b$  
in place of
$D,D_s,\eta _c.$ In this case, introducing $y>0$ through the relation
\beq
\alpha ^{'}_{b\bar{n}}=\alpha ^{'}_{b\bar{s}}=\frac{\alpha ^{'}}{1+y},
\eeq
one finds
\beq
\alpha ^{'}_{b\bar{b}}=\frac{\alpha ^{'}}{1+2y},
\eeq
and
\beq
(1+y)\left( \eta ^2+\eta _s^2\right) +\frac{2(1+y)}{1+2y}\eta  
_b^2=2\left( B^2
+B_s^2\right) .
\eeq
It follows from Eqs. (15),(18) for the measured vector meson masses, and 
$\alpha ^{'}_{c\bar{c}}\simeq 0.5$ GeV$^{-2}$ that $\alpha  
^{'}_{b\bar{b}}
\cong 0.182$ GeV$^{-2},$ consistent with the value $\alpha  
^{'}_{b\bar{b}}=1/
5.82\cong 0.172$ GeV$^{-2}$ found in ref. \cite{CN1}. This value of  
$\alpha ^{
'}_{b\bar{b}}$ supports the value
\beq
y\simeq 1.85.
\eeq
With this $y,$ one further obtains from (34) a relation similar to (31),
\beq
17.1\;K^2+3.64\;\eta _b^2=6\left( B^2+B_s^2\right) .
\eeq

Thus, the new sixth and 14th order mass relations may be accurately  
reduced to
quadratic ones by use of specific values for the Regge slopes.

For vector mesons, the values of the l.h.s. and r.h.s. of the  
formula (36)
are, respectively, (in GeV$^2)$ 339.4 vs. 346.1; the accuracy is  
therefore
$\sim 2$\%.

This formula (as well as Eqs. (19) and (22)) may be applied, for
example, to the question of the correct $q\bar{q}$ assignments for  
the meson
multiplets. There are two states in the most recent Review of  
Particle Physics
\cite{data} discovered recently whose quantum numbers are  
uncertain, the $B_J^
\ast (5732)$ and $B_{sJ}^\ast (5850)$ having masses $5698\pm 12$ MeV and
$5853\pm 15$ MeV, respectively. Since the dominant decay modes of  
the $B_J^\ast
(5732)$ are $B^\ast \pi $ and $B\pi $ \cite{data}, a natural  
suspicion would
be that this state is a tensor meson. Moreover, the relative  
proximity of the
masses of these states would suggest that the both belong to a common 
multiplet, i.e., to the tensor meson 25-plet. This suspicion may be  
tested in
two ways:

1) one may use the relation, as follows from the assumption on the
(quasi)-linear $B^\ast $ and $B_s^\ast $ Regge trajectories (with equal 
slopes) on which the discussed mesons lie,
\beq
m^2(B_2^\ast )-m^2(B^\ast )\simeq m^2(B_{s2}^\ast )-m^2(B_s^\ast ),
\eeq
which in this case gives (in GeV$^2)$ $4.11\pm 0.14$ vs. $4.92\pm  
0.18,$ or

2) one may use the formulas (19) and (36). For the values (as those  
for the
masses of the $B_J^\ast $ and $B_{sJ}^\ast ,$ respectively) $B=5.7$  
GeV and
$B_s=5.85$ GeV, the values of the l.h.s. and r.h.s., respectively,  
of Eq. (19)
are (in GeV$^2)$ 129.1 vs. 124.7, with an accuracy of $\sim 3.5$\%.  
For the
same values of $B$ and $B_s,$ the formula (36) yields (in GeV$^2)$  
418.7 vs.
400.3, with a similar accuracy of $\sim 4.5$\%, which does not  
exceed a limit
of $\sim 5$\% expected from the assumption on equality of the Regge  
slopes in
the light quark sector.

Finally, we discuss the failure of group theory to produce correct mass 
relations for heavy quarkonia. Consider, along with Iwao \cite{Iwao}, 
sub-$SU(3)$ symmetry which incorporates the $u$-, $d$- and  
$c$-quarks, i.e.,
assume that a world consists of the $u$-, $d$- and $c$-quarks, and the 
$c$-quark plays the same role as $s$-quark in the real world (the  
difference
in the masses of the $s$- and $c$-quarks is not reflected in a group 
theoretical approach). Then charm $(C)$ will play the role of  
strangeness
$(S),$ the ``supercharge'' $Z$ may be introduced in place of  
hypercharge,
viz., $(B$ stands for baryon number) $$Z=B+C,$$ the modified
Gell-Mann--Nishijima formula will become $$Q=I_3+\frac{Z}{2},$$ and the 
Gell-Mann--Okubo mass formula will take on the form
\beq
m^\ell =a+bZ+c\left[ \frac{Z^2}{4}-I(I+1)\right] ,
\eeq
where $\ell =1$ or 2, and $a,b,c$ are independent of $I$ and $Z$ but, in 
general, depend on $(p,q),$ where $(p,q)$ is any irreducible  
representation of
$SU(3).$ The whole difference of the above formula from the standard 
Gell-Mann--Okubo one \cite{GMO} is the presence of $Z$ in place of  
$Y$ in the
standard formula. Then, in the same way Eq. (26) follows from the  
standard
Gell-Mann--Okubo formula, the following relation may be derived  
from (38)
\cite{DMP}: $$\eta ^\ell +\eta _c^\ell =2D^\ell ,$$ which for $\ell =2$ 
disagrees with experiment, and for $\ell =1,$ although perhaps  
justified for
vector mesons, fails for other multiplets, as discussed in the  
beginning of
the paper. Thus, the failure of group theory to produce correct  
mass relations
for heavy quarkonia is solely due to the large bare masses of the  
$c$- and
$b$-quarks as compared to those of $u$-, $d$- and $s$-quarks. On  
the other
hand, group theory is rather successful in producing a correct mass  
formula in
the $(u,d,s)$-sector, since the difference in the masses of the  
non-strange
and strange quarks (and symmetry breaking in sub-$SU(3)$ sector of full 
$SU(4)$ or $SU(5))$ is negligibly small as compared to that of the  
strange and
charm quarks. The difference in the quark masses seems to be the  
only reason
for the different slopes of the corresponding trajectories. We also  
remark
that we do not see an alternative way to Regge phenomenology to  
obtain mass
relations for heavy quarkonia, at least at present.

In conclusion, we note that the derived Regge slopes in the charm  
sector are
\cite{su41}
\beq
\alpha ^{'}_{c\bar{n}}\simeq \alpha ^{'}_{c\bar{s}}\simeq  
0.63\;{\rm GeV}^{-2},
\;\;\;\alpha ^{'}_{c\bar{c}}\simeq 0.50\;{\rm GeV}^{-2},
\eeq
and in the beauty sector
\beq
\alpha ^{'}_{b\bar{n}}\simeq \alpha ^{'}_{b\bar{s}}\simeq  
0.30\;{\rm GeV}^{-2},
\;\;\;\alpha ^{'}_{b\bar{c}}\simeq 0.27\;{\rm  
GeV}^{-2},\;\;\;\alpha ^{'}_{b
\bar{b}}\simeq 0.18\;{\rm GeV}^{-2}.
\eeq

\end{document}